\documentclass[12pt,twocolumn]{IEEEtran}
\usepackage[T1]{fontenc} 
\usepackage{amsmath}
\usepackage[cmintegrals]{newtxmath}
\usepackage{bm} 

\usepackage{cite}

\usepackage{cleveref}
\usepackage{ dsfont }
\usepackage{bm}
\usepackage{amsmath,amssymb,epsfig,epsf, multirow,wrapfig, graphicx}
\usepackage{epstopdf}

\title{CogRF: A New Frontier for Machine Learning and Artificial Intelligence for 6G RF Systems}

\author{\IEEEauthorblockN{Tarun Cousik, Rubayet Shafin, Zhou Zhou, Kaleb Kleine, Jeffrey Reed, and Lingjia Liu} 
\thanks{The authors are with Wireless@Virginia Tech, Bradley Department of Electrical and Computer Engineering, Virginia Tech, Blacksburg, VA, USA.}}
\begin{document}	
\maketitle
	
\begin{abstract}
The concept of CogRF, a novel tunable radio frequency (RF) frontend that uses artificial intelligence (AI) to meet mission requirements for beyond 5G and 6G systems, is introduced. 
CogRF utilizes AI as the core to control and operate RF system components with the objective of optimizing the overall system performance. 
An overview of the vital elements that make up CogRF as well as the overall hierarchy of the envisioned CogRF system is provided, and potential RF components and control parameters are discussed. 
AI-powered flexible RF front ends, provide new opportunities to identify to enhance security, speed up optimization of device configurations, further refine radio design, improve existing spectrum sharing operations, and develop device health analytics. 
Top research challenges for CogRF systems have also been described and potential research directions are provided.
\end{abstract}

\section{Introduction}
Cooper's law \cite{cooper2003antennas} suggests the data rate available to wireless devices doubles roughly every thirty months.
Meanwhile, the advent of 5G enables new data-intensive and/or low-latency applications such as remote surgery, virtual reality, and online gaming. 
To meet this unprecedented surge in data demand, new technologies are needed for Beyond 5G and future 6G cellular networks. 
Specifically, significant improvements are needed in the radio frequency (RF) signal chain of the physical (PHY) layer, where conventional design techniques are being pushed to the limits where innovation is the only way forward. 
Traditionally, RF front ends are designed to meet custom specifications such as specific ranges of operational frequencies or radiation power levels. 
This design philosophy while reducing the complexity, renders the system inflexible to the adapt to the changing environment conditions. 
Next generation RF front ends should be able to modify their performance characteristics such as operational frequencies, RF level plans, non-linearity levels etc. 
Conventional tuning systems with digital and analog controllers suffer from two main drawbacks -- the first is that optimization typically takes place at the component's level. With the advent of fully digital RF architectures, it is important to recognize that the RF signal chain's complexity will significantly increase given that each analog-to-digital converter (ADC)/digital-to-analog converter (DAC) now connects to its own chain of power amplifiers (PAs), phase shifters, antennas, filters etc.  Adding additional component optimization modules to these systems further adds to the design complexity.
The second drawback, stems from challenges in implementation; traditionally, look-up tables (LUT) realize a controlled set of states in the RF systems. For example, a beamformer system may have sets of fixed weights it calls upon to point the beam in certain directions. 
While LUTs are convenient to implement in real time applications, the disadvantage rises from the limited number of configuration options. 
Additionally, there is no guarantee that the optimal solution for the current requirement is present in the LUT. 
For instance, when an unknown bearing angle which is not in the LUT is required, a beam former operating using a LUT will produce sub-optimal results irrespective of how carefully the LUT is designed. On the other hand, calculating exact values to be provided to various components using traditional computational techniques is computationally demanding.

Recent advances in artificial intelligence (AI) have triggered the adoption of machine learning (ML) tools into wireless systems\cite{Rawat2010,shafin2019artificial,patnaik2007ann,youssef2018machine,ciminski2005neural}.
Neural networks (NNs) essentially adjust themselves to perform a suitable set of mathematical functions that transforms the inputs to the outputs. When fully trained, if any unknown data set is provided at the input, it can correctly predict the output to a high degree of accuracy. Additionally, they are maybe less computationally expensive compared to the conventional algorithms. Therefore, ML offers a solid middle ground that is both computationally relaxed and relatively accurate.

In this work, we introduce \textit{CogRF}: a nascent idea that combines advances in AI and ML with RF to disrupt traditional radio design philosophies. The name stems from the idea that add the power of cognition (Cog) to RF front ends. In Section II, we elaborate the concept of CogRF, then develop a novel approach in designing its architecture, and highlight some of its functionality. In Section III, we investigate how the CogRF engine can optimally control commonly used RF components. In Section IV, we introduce some of the potential use cases of CogRF systems. Section V highlights some research challenges CogRF systems face moving forward. Section VI provides an executive summary of the key concepts covered in this work.
\section{CogRF-Concept and Architecture}
CogRF is defined as \emph{RF architectures that utilize AI to realize optimal configurations for their RF components, obtain better awareness of the RF environment, or acquire the ability to predict RF events through reasoning and learning.} 
The learning ability of the CogRF system emboldens the RF signal chain with the power of generalization and inference. CogRF systems can learn optimal RF component configurations from previously sensed environment data and tune control knobs of RF components to react to real-time data in an 'intelligent' fashion thereby boosting system performance. One popular approach begins with teaching the NN to dynamically engage with an environment similar to the operational environment, as is the case of any reinforcement learning type of framework. Subsequently, when required, real time data can be utilized alongside the training set data to help develop systems which optimize themselves to the operational environment; thereby producing systems that improve with time. This data-driven learning based approach equips CogRF systems with a high amount of immunity against any non-idealities or deviation from a presumed model.
By the property of self-awareness, CogRF systems are cognizant of the characteristics of the RF front end. Therefore, CogRF systems are aware of hardware malfunction/changes on any element of the RF chain. This enables CogRF systems to dynamically configure themselves to next best RF parameter settings after taking into account the effects of faulty component. 
\begin{figure}
\centering
\includegraphics[width=0.9\linewidth]{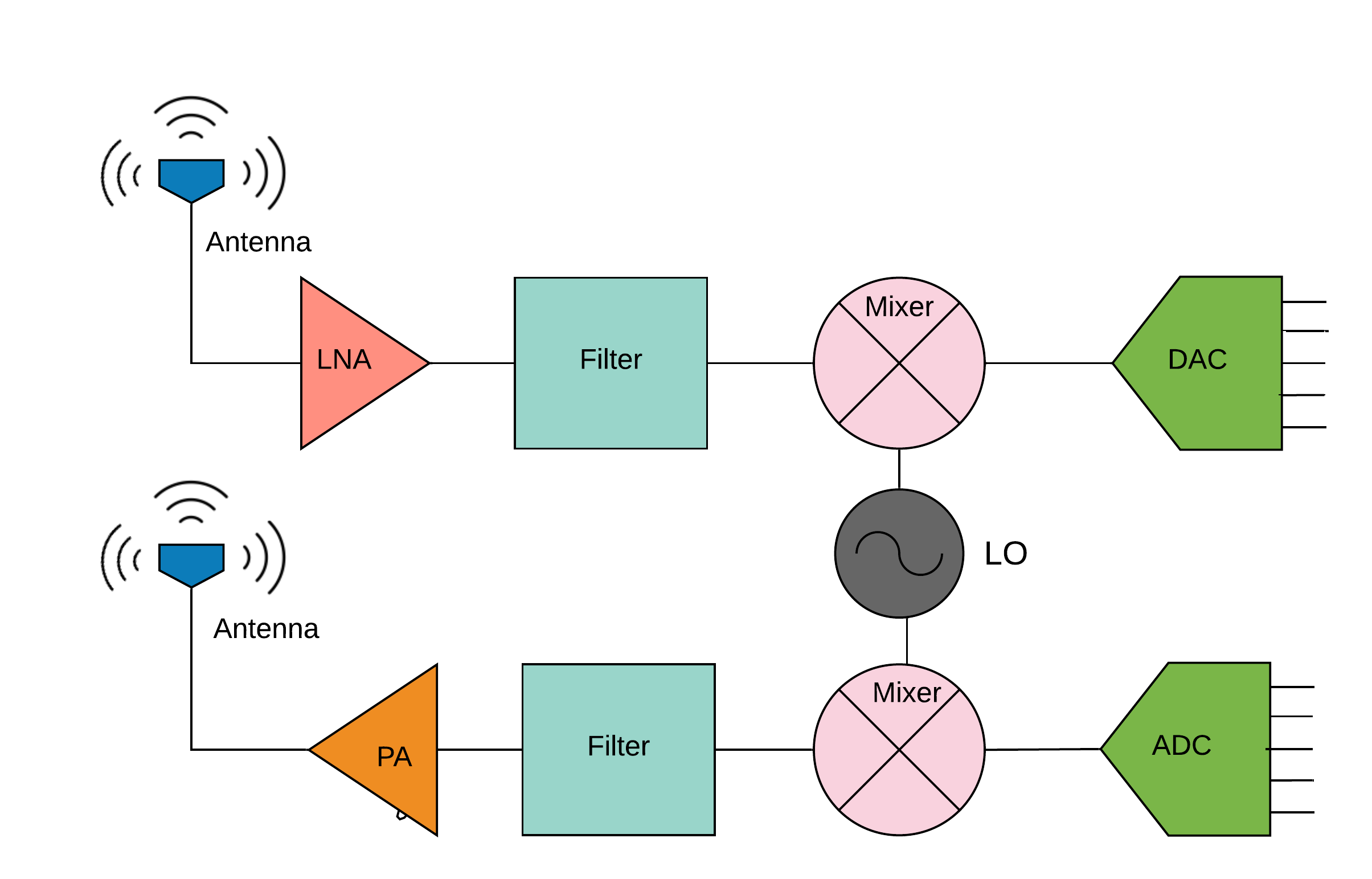}
\caption{Vanilla RF Transmitter and Receiver Chains}
\label{Vanilla_TxRx}
\end{figure}

This work is fundamentally different from previous Cognitive Radio (CR) work that dealt with topics like  modulation classification, interference detection, device identification and authentication, beam management etc--- processes that are implemented in the digital signal processing layer or higher.
CogRF specifically deals with utilizing AI/ML in optimizing the RF front end thereby filling up the missing piece of the puzzle that CR left out. 

\textbf{CogRF Architecture and Hierarchy:}
\begin{figure*}
\centering
\includegraphics[width=1\linewidth]{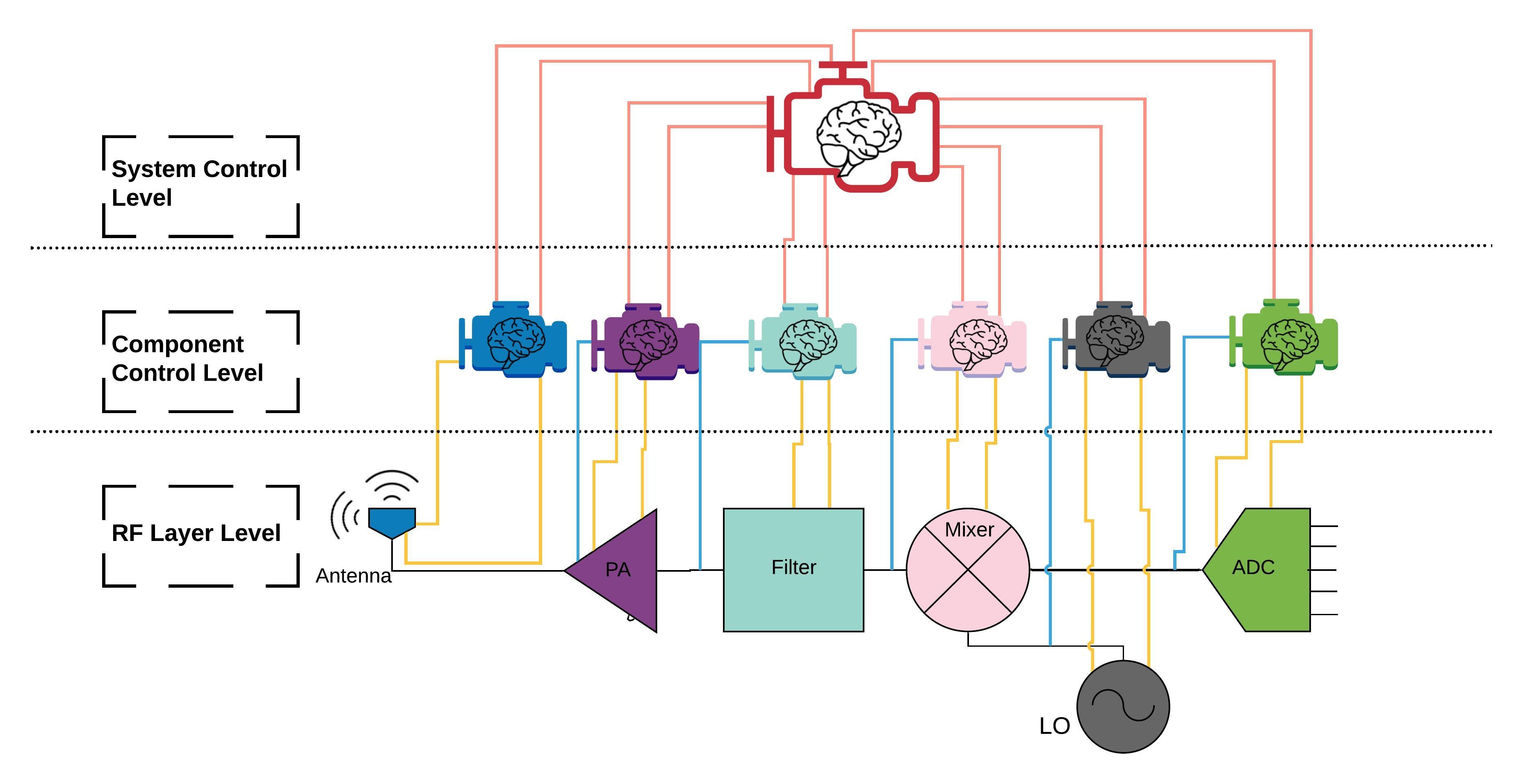}
\caption{Illustrating the CogRF Transmitter .}
\label{CogRF_system}
\end{figure*}
Fig.~\ref{CogRF_system} shows the CogRF architecture:

\textit{AI/Cognitive Engine} (represented as the brain in the engine): This is the metaphorical brain of the CogRF system which utilizes data sensed from various RF components and uses it alongside prior experience and/or device models to suggest appropriate RF parameter settings to optimize performance. 

\textit{Sensor lines} (blue lines):The sensor lines provide the Cog Engine a means to continuously monitor the RF chain. Typical sensory data the Cog Engine would find useful include spectrum occupancy, incoming power, non-linearity levels, detection of abnormal signals such as those of interference or those characteristic of faulty equipment etc.

\textit{Control lines} (yellow lines): Once the AI engine identifies the optimal parameter set, RF controller lines are used to apply the new parameters to the respective RF components.

The hierarchy of the CogRF system is also visually represented in Fig. 2. It comprises of three layers with the following function:

\textit{The RF Layer:}
RF layer comprises of the vanilla RF front end with additional control and sensory lines added. In a transmitter, this would include everything between the DAC to the antenna while in the receiver it would involve everything between the antenna to the ADC. For the sake of brevity,  Fig. 2 shows only the CogRF Transmitter.  
 
\textit{The Component Optimization Layer:}
This layer consists of multiple specialized component level NNs (CLNs) that are trained to interface and control a specific component. In Fig. 2, the specialized NNs and the components they control are represented with the same color. The NNs in this layer are pre-trained to map various input/output responses of their respective components. The idea of assigning a NN to each component simplifies the complexity of the NNs required to control and configure various components. 
The sensory line ensures that the NN is constantly updated with information on the component's performance. The control lines allow the NN to dynamically change operating parameters of the component under its control. 
 
\textit{The System Optimization Layer:}
The system optimization layer consists of a System Level NN (SLN) whose sole purpose is to ensure system level optimization to achieve the mission requirement of the transceiver. The SLN achieves this by dictating high level component performance requirements to each CLN. The high level requirements are then processed by the CLNs to determine the configuration of their components using the control lines. 

\textit{An Example:}
The SLN may determine that a required signal-to-noise-ratio (SNR) of $20$ dB. It will collect current component performance status from the CLNs, make a decision on how to achieve this $20$ dB requirement. Based on the determination, it will then dictate high level requirements to the CLNs. In this case, this may include instructions such as increasing the PA gain by $10$ dB and changing the bearing of the antenna array pattern to point to a new angle of $30$. The CLNs for the PA and the array receive their respective requirements and start translating this into their requirements. The CLN controlling the PA, would then calculate the increase in bias current needed by the PA to dictate said change and inform the PA to do so. For the CLN controlling the array, this includes calculating the new phase weights each element needs to be assigned and then providing that information to the phase shifters. Note that while this hierarchical architecture may enable optimal operation of the entire system, it is not necessary for any or all individual components be operating optimally. 
\textbf{RF Configurable Parameters and CogRF Functionalities:} CogRF functionalities in transmitter and receiver chain are different. 
At the transmitter, main objectives are to minimize the power consumption or to restrict transmission within the band of interest to meet adjacent channel level ratio requirements. Accordingly, CogRF can adapt to varying transmission requirements and waveform characteristics. Meanwhile, the objective at the receiver is to maximize SNR at the output of the ADC. CogRF is able to control the corresponding parameters to fulfill the receiver requirements.
A summary of the CogRF controllable parameters for different RF components is shown in table \ref{table:RFparameters}.
	\begin{table}[ht]
\caption{Example RF Controllable Parameters}
    \centering
\begin{tabular}{c c}
\hline\hline
RF Component  & Tunable Parameters \\ [0.5ex] 
\hline
DAC/ADC   & Sampling rate, number of bits per sample \\
\hline
LNA & Gain, current bias \\
\hline
LO & Oscillator frequency, drive level\\
\hline
Filter & Selectivity, operational frequency, bandwidth\\
\hline
Power amplifier & Voltage and current bias, antenna loading\\
\hline
Antennas & Excitation amplitude, phase, matching\\
\hline
\end{tabular}
\label{table:RFparameters}
\end{table}

\section{Operational Perspective of CogRF}
This section discusses component performance optimization of the CogRF engine. 
\subsection{Arrays}
Cellular networks are facing challenges such as rapidly user movements, varying channel conditions and service strict compliance requirements. 
This is especially true with millimeter-wave systems where pencil beams are adopted. There are two potential areas where NNs can help boost array performance. 

(i) \textit{Beam-Shaping and Beam-Switching:}  Due to spectrum sensing and cognition capability, CogRF systems are able to create accurate beams for multi-user (MU) MIMO processing with low computational complexity. This precise beam-shaping will significantly reduce the interference to unintended users. For example, in \cite{zooghby1998neural}, Radial Basis Functions (RBFs) are used to find phase and amplitude solutions for a $10 \times 10$ array that would simultaneously serve $9$ users while nulling $10$ spatially distinct interferers. 
Furthermore, if considered the possibility of receiving user feedback, the CogRF engine can additionally enable fast and optimal beam-switching to adapt with channel variation or user mobility.
	
(ii) \textit{Array Calibration:} 
Typically array calibration rectifies the far field (FF) patterns, minimizes mutual coupling between elements, reduces side lobe levels, and identifies anomalous antenna elements in the array. Autonomous calibration is crucial for Beyond-5G or 6G antenna arrays and this is especially important for massive MIMO or full-dimensional MIMO systems. The NNs within CogRF's inference engine can be trained to self-calibrate. When deployed real-time, these systems can optimize themselves to minimize mutual coupling or optimize the FF patterns. NN-based self-calibration approach with lower computational complexity have shown great promise in outperforming the traditional array calibration methods \cite{long2014calibration, patnaik2007ann, host2018digital}. 

\subsection{Power Amplifiers (PAs)}
CogRF can serve a variety of different support roles when integrated with PAs. NNs, when implemented alongside PAs, can be trained to reduce power consumption, adjust gain settings, change the output power as required or in generating coefficients for pre-distortion. When modeling the learning system, the bias currents and voltages can be considered as the inputs to the NN. The corresponding power consumed and PA gain can serve as the outputs of the underlying NN.  
        
For energy efficiency, PAs are often operated in the non-linear regime, and different pre-distortion techniques are utilized for linearizing the output of the PA.
Analog pre-distortion can be realized using NNs which optimize the control coefficients \cite{ciminski2005neural}. 
During the training phase, NN will learn the  mapping for the input-output system response. 
During the online operation, upon sensing the power spectrum, NN within the CogRF engine will suggest the best operating regime that minimizes the intermodulation distortion. 
Whenever the performance falls below a threshold, the CogRF engine will be triggered to retrain the NN with the most recent data. 
NNs can also be trained for digital pre-distortion for potential improvement in the linearization of the AM/AM effect and reduction in the peak-to-average-power-ratio~\cite{Rawat2010}.
	
\subsection{ Filters }
Ideal RF systems should be able to nimbly hop between multiple separated frequency bands and operate over different bandwidths. Being able to intelligently tune filter's to meet the various frequencies as well as bandwidths in a time sensitive manner is a considerable challenge. NNs can provide this time critical intelligence required to tune the filters when they learn to recognize the best filter configurations. 
In~\cite{michalski2010artificial}, NNs are trained to optimize filter S-parameters using tunable screws. 
Starting with an already tuned filter, screw positions for the tuned filter are recorded to generate a mapping between the reference screw position and filter's ideal transfer function. 
When in operation, CogRF engine identifies the non-ideal screw position based on the measured S-parameters and rotates the screw to tune the filter back to the optimal position. This method can be extended to MEMs filters to allow fast and intelligent frequency tuning of realistic small filters using electrical signals.

\subsection{ADC}
The CogRF engine, when integrated with the ADC, can help in actively managing power consumption. The ADC's bit-resolution dictates the signal-to-quantization-plus-noise ratio (SQNR) at the output of the ADCs~\cite{le2005analog}; increasing the number of bits improves the SQNR, usually at the cost of higher power consumption. The CogRF engine can be trained to optimize this trade-off: over time it can learn to adjust the bit resolution to maintain a threshold SQNR while minimizing power consumption. In this scenario, the SQNR values would be fed to the CogRF engine through the sensory lines and the engine would appropriately step up or down the resolution to meet requirements to optimize power consumption based on the training it has received. 
		
Another important aspect of ADCs is in filtering.  Sampling rate of the ADC can be intelligently utilized to reject signals that fall beyond the first Nyquist zone of the ADCs. Sampling frequency of the ADCs can be adjusted based on the observation of the power levels and nature of the neighboring signal so that aliases from higher order Nyquist zone do not fold into the desired channel.
		
\section{Use Cases:}
\subsection{Use Case 1: MIMO Self-tuning Sectorization}
Apart from the user-specific MIMO  beamforming, cellular networks also need to perform sector-specific beamforming for network coverage enhancement or broadcast operations. Presently, RF parameters corresponding to sector-specific broadcast beamforming are manually set based on drive-test results, and kept unchanged for a long period of time. These fixed RF parameters do not account for changes in environment or users' mobility patterns resulting in a strictly sub-optimal solution.
 
\begin{figure}
    \centering
    \includegraphics[width=1.0\linewidth]{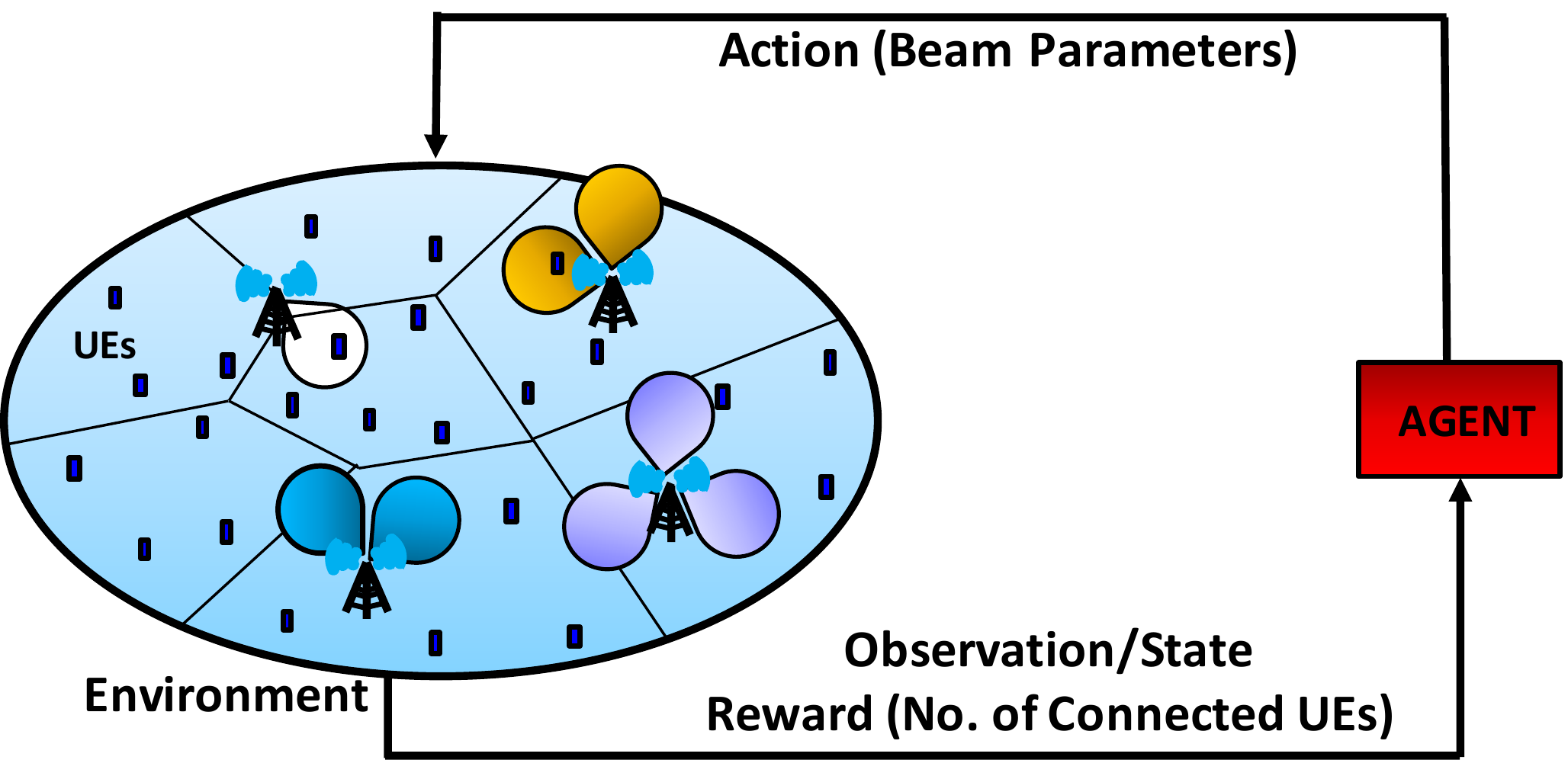}
    \caption{DRL-based CogRF beam learning framework.}
    \label{framework}
\end{figure}

To maximize the network coverage, RF parameters controlling sectorization need to be dynamically updated depending on changes in users' mobility.
In \cite{shafin2019self}, a deep reinforcement learning (DRL) based approach is utilized to automatically update the RF parameters based on changes in users' mobility. 
The CogRF's AI engine can host the DRL agent for controlling the RF parameters through interactions with the changing environment. 
The CogRF AI engine observes the state of the network through the sensor systems, and takes action (in this case, a particular setting of RF parameters) that maximizes the reward or network performance metric. 
The interplay between state, action, and reward is shown in Fig.~\ref{framework} where the CogRF engine can dynamically adapt with users' mobility pattern, and select the best RF beam parameters.
\subsection{Use Case 2: Federated Learning for Wireless}
Federated learning is becoming a mainstream machine learning approach enabling deployment of accurate generalized models across multiple devices. 
It consists of one central node (typically a server) and several slave nodes (in our case, CogRF systems) serviced by the central node. 
The slave nodes sense the environment and optimize themselves to prevailing environmental conditions. 
The slave nodes then proceed to send the central node any insights they gained such as the configurations they've found to be optimal for various environmental conditions, overall system health, component health, incidents of malfunctions or link loss.
In federated learning approaches the central node aggregates and generalizes these insights and updates all the slave nodes. 
The CogRF system provides a natural platform for adopting federated learning for wireless systems.
To be specific, we envision that generalized component configuration models can be developed for various deployment scenarios. 
In this way, incoming users who just enter the service area can quickly configure themselves to a `close to optimal' state from where they can further optimize themselves based on federated learning. 
The feedback received about the health of the systems and components promises to help designers and maintenance personnel address component failures before they happen, choose appropriate components for the future designs as well as allow for graceful failure and design systems which gain additional reliability through peer learning. 
Future designers can routinely check up on millions of devices and send out software patches which provide work-arounds to potential RF failures.  

\subsection{Use Case 3: Security}
The increase in public scrutiny and demand for user privacy, the advent of health devices connected to wireless networks and the rise in sophisticated data siphoning techniques are key reasons to look for new ways to enhance security measures. CogRF offers a shot at enhancing conventional security measures used in wireless systems at little to no extra cost. Given that the CogRF engine constantly monitors its components to changes in the environment, it is very aware of the flaws and implementation related non-idealities of its individual components. These flaws are a combination of variation in manufacturing, deployment conditions  as well as operational fatigue. Impairments such as IQ imbalances, frequency and DC offsets, as well as non linearities from the components can be utilized to provide a unique fingerprint/identity to each RF device \cite{kim2008specific, youssef2018machine}. Combined with a Federated Learning approach, it provides a convenient path to share the individual quirks of each system thus enabling an easy way to authenticate and identify devices.

\subsection{Use Case 4: Spectrum Sharing }
CogRF systems can provide two unique opportunities when it comes to spectrum sharing.
First, when CogRF systems are connected to a central server (such as the one in Use Case 2), one can provide the server with the ability to act as a spectrum access system (SAS). A SAS dictates when, where, and to whom the spectrum is to be allocated, and this information could then be easily dispersed onto all the slave nodes. This approach allows each level of the spectrum sharing hierarchy to have a manageable amount of computational overhead while ensuring nimble operations. 

Second, spurs arising from RF non-linearities can be falsely detected as signals from other systems operating in the band. 
This false detection entails two detrimental consequences -- systems may decide the spectrum is occupied and wait; thereby reducing the spectrum usage efficiency or they may transmit at higher power levels; which leads to inefficient use of power. 
Both these detrimental consequences have serious financial costs which can simply be avoided with the help of CogRF systems: CogRF systems can be trained to push out the spurs from the frequency bands of interest by re-configuring their mixers or ADCs.

\section{Top Challenges}

\subsection{Limited Training Data} The key ingredient behind the success of artificial intelligence or machine learning is data-driven approach for training. For effectively training CogRF systems, we need a significant amount of training data. For example, in order to train a CogRF engine responsible for learning the optimal sampling rate and the number of bits per samples, the learning engine needs to experience the prior relevant data. There are two sets of data that need to be considered for training: (i) sensory data collected by the sensing lines (ii) Prior Action Data (PAD). PAD pertains to the parameter setups which resulted in a particular desired performance. For example, when selecting ADC sampling rate, the AI engine will need the information on what prior sampling rate resulted in particular SINR value at the output of the ADC under similar environmental sensing condition. This data intensive adaptation process creates two major research challenges-- 1) How do we obtain the large and diverse sets of data required for effectively training the CogRF systems? 2) How do we train the CogRF systems with fewest sets of training data? 

\subsection{Algorithm Design for RF Training}	
There is a significant push for designing ML algorithms and techniques for wireless systems \cite{jiang2016machine}. 
A key takeaway from \cite{jiang2016machine} is that ML algorithms often need to be tailored towards the particular problems at hand. Therefore, this should be carefully considered when designing algorithms to train CogRF engines which control and coordinate between diverse components of the RF chain. Additionally, training must also consider the physical constraints and characteristics of respective component.

\subsection{Power Consumption and Energy Efficiency}
Large amount of data needs to be processed when training NNs. Hardware implementation of AI algorithms must take into account the aspects of computation -- low complexity/fast processing time and high energy efficiency. 
Faster processing, and hence faster convergence of any machine learning training, is important for applications that require prompt adaptation or prediction. 
On the other hand, designing solutions with low power consumption for data-intensive computations is another challenging aspect for researchers.
Neuromorphic computing (NC) has been shown to be extremely energy efficient with the analog version being $4$ times more efficient than its digital counterpart~\cite{YI2016175}; hence, by integrating NCs with CogRF it might be possible for CogRF systems to have much better power consumption than conventional RF systems even with the need for power to support the AI.
Future research needs to be conducted on energy efficient training algorithms tailored towards RF systems as well as in designing efficient hardware implementations.
\section{Conclusion}
CogRF systems represent a sharp paradigm shift to achieve system level optimization with the aid of AI. The intelligence and capability to adapt is distributed over three layers. The highest layer comprises of a NN called the system level network which is trained to dictate appropriate high level requirements to the middle layer as and when required. The middle layer consists of a set of unique NNs that interface with individual RF components of the lowest layer. The NNs in the middle layer are trained to understand how to configure the RF components to meet the high level requirements dictated by the system Level network. The lowest layer comprises of the RF layer which consists of highly customizable RF front ends which execute their functions. CogRF opens new avenues to  manage device health diagnostics, provide better cellular infrastructure, improve spectrum sharing strategies, and aid in providing  device security.

\section*{Acknowledgements}
The authors thank the U.S. National Science Foundation (NSF) for funding this work under grants CNS-1642873 ,ECCS-1802710, ECCS-1811497, CNS-1564148 and CNS-1811720. 
\bibliography{CogRF}

\begin{thebibliography}{10}
\providecommand{\url}[1]{#1}
\csname url@samestyle\endcsname
\providecommand{\newblock}{\relax}
\providecommand{\bibinfo}[2]{#2}
\providecommand{\BIBentrySTDinterwordspacing}{\spaceskip=0pt\relax}
\providecommand{\BIBentryALTinterwordstretchfactor}{4}
\providecommand{\BIBentryALTinterwordspacing}{\spaceskip=\fontdimen2\font plus
\BIBentryALTinterwordstretchfactor\fontdimen3\font minus
  \fontdimen4\font\relax}
\providecommand{\BIBforeignlanguage}[2]{{%
\expandafter\ifx\csname l@#1\endcsname\relax
\typeout{** WARNING: IEEEtran.bst: No hyphenation pattern has been}%
\typeout{** loaded for the language `#1'. Using the pattern for}%
\typeout{** the default language instead.}%
\else
\language=\csname l@#1\endcsname
\fi
#2}}
\providecommand{\BIBdecl}{\relax}
\BIBdecl

\bibitem{cooper2003antennas}
M.~Cooper, ``Antennas get smart,'' \emph{Scientific American}, vol. 289, no.~1,
  pp. 48--55, 2003.

\bibitem{Rawat2010}
M.~{Rawat}, K.~{Rawat}, and F.~M. {Ghannouchi}, ``Adaptive digital
  predistortion of wireless power amplifiers/transmitters using dynamic
  real-valued focused time-delay line neural networks,'' \emph{IEEE Trans.
  Microw Theory Techn.}, vol.~58, no.~1, pp. 95--104, Jan 2010.

\bibitem{shafin2019artificial}
R.~Shafin, L.~Liu, V.~Chandrasekhar, H.~Chen, J.~Reed, and J.~Zhang,
  ``{Artificial Intelligence-Enabled Cellular Networks: A Critical Path to
  Beyond-5G and 6G},'' \emph{preprint arXiv:1907.07862}, 2019.

\bibitem{patnaik2007ann}
A.~Patnaik, B.~Choudhury, P.~Pradhan, R.~Mishra, and C.~Christodoulou, ``An ann
  application for fault finding in antenna arrays,'' \emph{IEEE Trans. Antennas
  Propag.}, vol.~55, no.~3, pp. 775--777, 2007.

\bibitem{youssef2018machine}
K.~Youssef, L.~Bouchard, K.~Haigh, J.~Silovsky, B.~Thapa, and C.~Vander~Valk,
  ``Machine learning approach to rf transmitter identification,'' \emph{IEEE J.
  Radio Freq. Identif.}, vol.~2, no.~4, pp. 197--205, 2018.

\bibitem{ciminski2005neural}
A.~S. Ciminski, ``Neural network based adaptable control method for
  linearization of high power amplifiers,'' \emph{AEU Int'l J. Electron. and
  Commun.}, vol.~59, no.~4, pp. 239--243, 2005.

\bibitem{zooghby1998neural}
A.~E. Zooghby, C.~Christodoulou, and M.~Georgiopoulos, ``Neural network-based
  adaptive beamforming for one-and two-dimensional antenna arrays,'' \emph{IEEE
  Trans. Antennas Propag.}, vol.~46, no.~12, pp. 1891--1893, 1998.

\bibitem{long2014calibration}
R.~Long, J.~Ouyang, F.~Yang, Y.~Li, K.~Zhang, and L.~Zhou, ``Calibration method
  of phased array based on near-field measurement system,'' in \emph{IEEE
  Antennas Propag. Society Intl. Symposium}.\hskip 1em plus 0.5em minus
  0.4em\relax IEEE, 2014, pp. 1161--1162.

\bibitem{host2018digital}
N.~Host and K.~O'Haver, ``Digital array planar near-field calibration using
  element plane wave spectra with iterative search,'' in \emph{US Nat'l.
  Committee of URSI Nat'l. Radio Sci. Meeting}.\hskip 1em plus 0.5em minus
  0.4em\relax IEEE, 2018, pp. 1--2.

\bibitem{michalski2010artificial}
J.~J. Michalski, ``Artificial neural networks approach in microwave filter
  tuning,'' \emph{Progress in Electromagn. Research}, vol.~13, pp. 173--188,
  2010.

\bibitem{le2005analog}
B.~Le, T.~W. Rondeau, J.~H. Reed, and C.~W. Bostian, ``Analog-to-digital
  converters,'' \emph{IEEE Signal Process. Mag.}, vol.~22, no.~6, pp. 69--77,
  2005.

\bibitem{shafin2019self}
R.~Shafin, H.~Chen, Y.~H. Nam, S.~Hur, J.~Park, J.~Zhang, J.~Reed, and L.~Liu,
  ``Self-tuning sectorization: Deep reinforcement learning meets broadcast beam
  optimization,'' \emph{preprint arXiv:1906.06021 [eess.SP]}, 2019.

\bibitem{kim2008specific}
K.~Kim, C.~M. Spooner, I.~Akbar, and J.~H. Reed, ``Specific emitter
  identification for cognitive radio with application to {IEEE} 802.11,'' in
  \emph{IEEE GLOBECOM}.\hskip 1em plus 0.5em minus 0.4em\relax IEEE, 2008, pp.
  1--5.

\bibitem{jiang2016machine}
C.~Jiang, H.~Zhang, Y.~Ren, Z.~Han, K.-C. Chen, and L.~Hanzo, ``Machine
  learning paradigms for next-generation wireless networks,'' \emph{IEEE
  Wireless Commun.}, vol.~24, no.~2, pp. 98--105, 2016.

\bibitem{YI2016175}
Y.~Yi, Y.~Liao, B.~Wang, X.~Fu, F.~Shen, H.~Hou, and L.~Liu, ``{FPGA} based
  spike-time dependent encoder and reservoir design in neuromorphic computing
  processors,'' \emph{Microprocessors and Microsystems}, vol.~46, pp. 175 --
  183, 2016.

\end{thebibliography}

\end{document}